# A Concept for a Scalable 2 kTon Liquid Argon TPC Detector for Astroparticle Physics

D.B. Cline[a] and F. Sergiampietri[a,b]

[a] *Department of Physics, UCLA, Los Angeles, CA 90024, USA*
[b] *INFN - Sezione di Pisa, Largo B. Pontecorvo, 3, 56127 PISA, Italy*

**Abstract**

*This paper describes the results of a study on the general lines, main construction criteria, crucial points, parameters and required preliminary R&D activities for the construction of a LAr (liquid argon) imaging detector with active mass in the 10-100 kTon range. Such detectors are crucial for supernova detection, proton decay, LBL neutrino physics and other astroparticle physics applications.*

## 1   Introduction

The on-surface test of the 300-ton ICARUS module made in 2001[1], demonstrated the 3d-imaging potentiality of this detection technique, only comparable to that of a bubble chamber but extended at many order of magnitude larger volumes. Due to its multi-module configuration and to its large liquid nitrogen consumption (~1 liquid $m^3$/hour), the 300-ton geometry and construction technique do not appear suitable for the design of a much larger mass detector based on this detection technique.

The preliminary study[2,3] on the modularity and the shape for a future large-scale detector, named LANNDD (Liquid Argon Neutrino and Nucleon Decay Detector), gave indication on the advantages of a single-module cylindrical cryostat. This conclusion is valid only if valid solutions are found for keeping the same construction/operation quality required for a detector based on an ultra high purity (UHP) liquefied noble gas and for coping with the engineering and safety issues related to the extended scale.

The huge costs, the multi-year construction and commissioning phase required for such a project can only be justified by a credible physics plan that should include natural and artificial neutrino oscillation physics and, seen the unique large instrumented mass combined with the low energy detection threshold, nucleon decay physics. This programme can only be performed by conceiving the detector as sited in an underground laboratory (in the USA, in Europe or in Japan) and in the beam line (in-axis or off-axis) of existing or future neutrino beams.

In the following we describe a further step on the study of possible realistic solutions and the required preliminary R&D activities for such a detector. We would like to underline that, in a large scale, the main difficulties do not arise from the detection technique, that even with some possible improvements or changes is the well-established ICARUS one, but from the engineering choices in the cryostat configuration affecting performances (purity levels, thermal insulation), mechanical stability, safety and the related construction and operation costs.

The road map to a final credible design should pass-through the project and construction of a reduced scale prototype, configured as scalable to the final mass. Due to its costs, such a prototype should be able to provide information not only on the optimized construction technique but also on its detection performances with cosmic particles and hopefully artificial neutrino's, by siting it, for example, on a ν–beam off-axis position. A detector with an active mass in the 2-4 kTon range seems the right choice for such purpose. The study and the remarks described in the following refer to a 2 kTon LAr (liquid argon) TPC detector.

## 2   Project guidelines

As shown in ref. 2, by optimizing the surface-to-volume ratio ($S/V$) heat input, wall out-gassing and number of wires are minimized, while the fiducial-to-active volume ratio is optimized. A cylindrical vessel with the height equal to the diameter has the same $S/V$ ratio than a spherical one inscribed in it or a cubic one surrounding it (see Figure 1). With the ionization read-out based on wire chambers (planar), the cubic shape appears the most convenient for the uniform drift times required to read the active volume; on the other side a pure cubical shape is not suitable to large sizes due to its mechanical instability. If, from one side the spherical shape requires half of the wall stiffness in comparison with the cylindrical one, the latter appears more suitable for the TPC drift geometry and with a more advantageous ratio between the detector volume and the occupied laboratory volume.

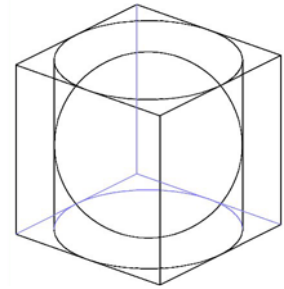

**Figure 1**

The previously mentioned physics program implies a running period of tenths of years. An optimized thermal insulation, based on double walls with vacuum in between and superinsulation layers wrapped around the inner vessel, appears as mandatory to limit the running costs at acceptable levels.

In a not negligible fraction [a], construction costs depend on the number of wires and associated electronic channels. A practicable way to minimize this number is to widen the drift region between each cathode and the facing wire chamber. An R&D activity [4] is under preparation at CERN for verifying the detector response at drift paths up to 5 meters. In the present study, we have assumed a cathode-to-wire chamber distance of 5 meters. Operation with such drift lengths sets serious constraints on the LAr purity (impurities ≤ 10-20 ppt $O_2$ equivalent). In order to reach and maintain during years such level of purity we consider as inalienable the following construction criteria and conditions:

*a)* the possibility of generating vacuum inside the inner vessel and of checking its tightness,

*b)* the use of stainless steel for the inner vessel walls [b] and for cathodes, for wire chamber frame and for electrical field shaping electrodes,

*c)* the continuous, adiabatic [c] argon purification in liquid phase,

*d)* the UHP and UHV (ultra high vacuum) standards for any device or cryogenic detail (flanges, valves, pipes, welding) in contact with the argon.

## 3  The cryostat

A cryostat with the following main parameters has been studied for a 2 kTon active mass detector (approximate values for higher detector masses are given for comparison):

| | | | | |
|---|---:|---:|---:|---:|
| **Active LAr** | | | | |
| mass [kTon] | **2** | 4 | 10 | 100 |
| diameter = height [m] | **12.2** | 15.5 | 20.9 | 45 |
| volume [$10^3 \cdot m^3$] | **1.43** | 2.9 | 7.2 | 71.6 |
| **Inner vessel** | | | | |
| diameter = height [m] | **13.0** | 16.5 | 22.3 | 48.0 |
| outer surface [$m^2$] | **797** | 1′591 | 2′906 | 13′464 |
| volume [$10^3 \cdot m^3$] | **1.7** | 3.5 | 8.7 | 86.9 |
| **Outer vessel** | | | | |
| diameter [m] | **15.0** | 19.0 | 26.0 | 55.5 |
| height [m] | **16.0** | 20.3 | 27.4 | 59.1 |
| Total heat input [d] [kW] | **1.6** | 3.2 | 5.8 | 27.0 |
| $LN_2$ consumption [$m^3$/day] | **1** | 2.1 | 3.7 | 17.3 |

As first choice, seen the heat input values, the thermal stabilization in thought as obtained through $LN_2$ heat exchangers.

---

[a]  For the ICARUS T600 detector, about 2/3 of the construction costs were proportional to the number of wires.

[b]  As possible compromise, the use of plate of a carbon steel or low-alloy steel base to which is integrally and continuously bonded on both sides a layer of stainless chromium steel (ASTM A263-03 Standard Specification for Stainless Chromium Steel-Clad Plate) can be considered for the inner vessel after positive tests on out-gassing and at low temperature.

[c]  By this term we indicates a purification system that circulates the liquid argon in large volumes at very low velocity through $O_2$-getter filters, such to not modify its thermodynamic conditions and to not generate an extra heat input.

[d]  Unitary heat inputs lower than 1 $W/m^2$ are reachable with vacuum and super-insulation layers. We assumed 2 $W/m^2$ to take into account residual conduction heat inputs through signal cables, mechanical supports and spacers.

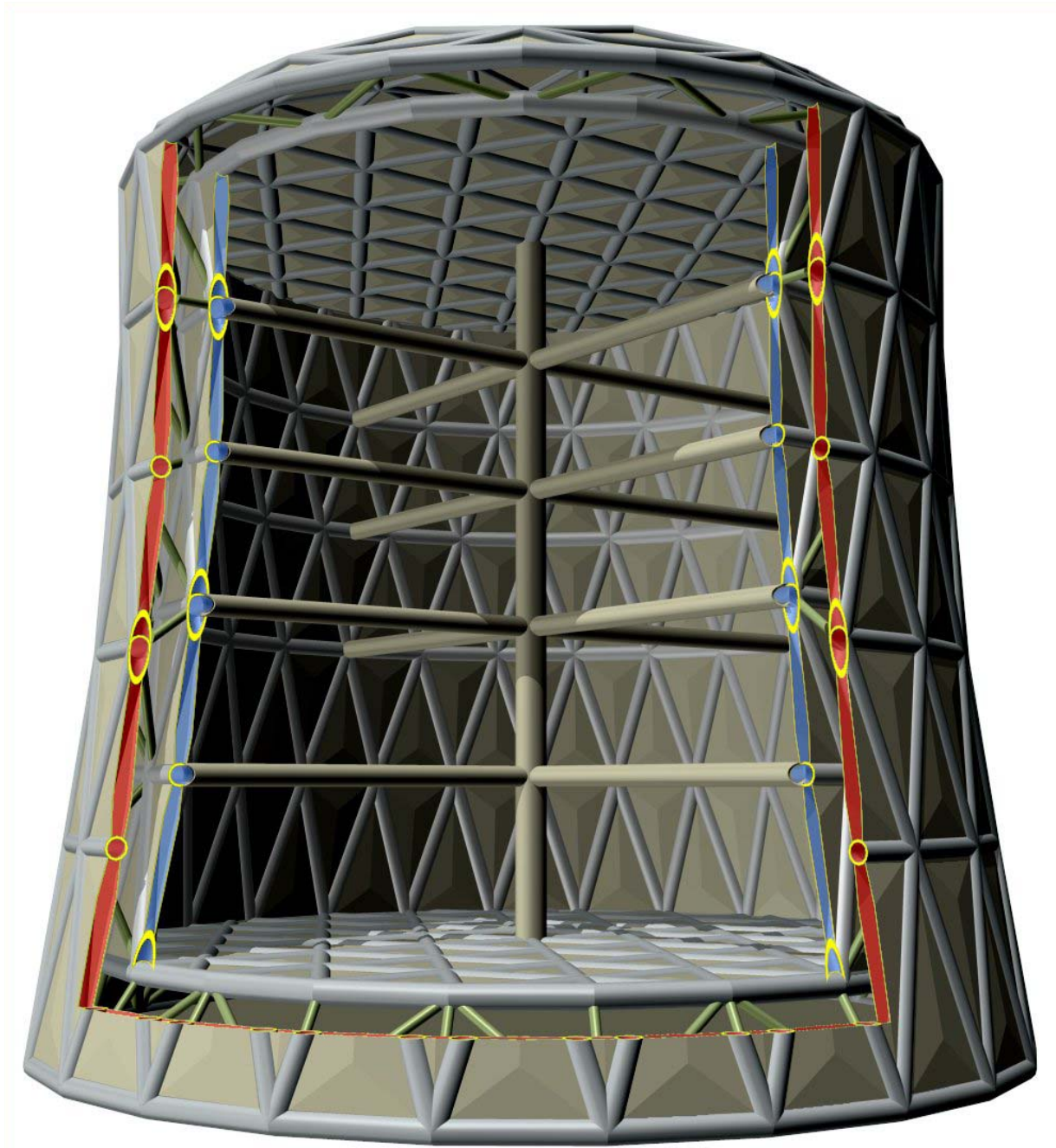

**Figure 2** - Cutaway view of the proposed cryostat design. The inner cross-structure is indicated for an eventual increase of the inner vessel rigidity.

The modest total heat input and $LN_2$ consumption reflect the stable and safe thermodynamic conditions in which the detector can be operated. Slow linear flows for nitrogen inside the heat exchanger and for argon through the purifier reduce at the minimum the microphonic noise that can affect the *S/N* ratio on the read-out wires. In case of cryogenerator-based thermal stabilization, the good thermal insulation reduces to negligible levels the electric power requirements.

Quiet and stable running of the detector is also a requirement to cope with the safety measures in an underground laboratory.

A valid project for the cryostat is matter for a technical study of the different geometries and solutions by a specialized engineers group. This should include structural optimization, FEA simulation and experimental tests at room and $LN_2$ temperatures. Here we will limit ourselves to underline the advantages and the difficulties for the overall detector operation of different possible general configurations. As working hypothesis, an interesting solution for the cryostat is described in Figure 1. Its design is based on a system of two vessels – cold inside and warm outside - each with walls composed by an array of SS (stainless steel) lattice girders whose triangular windows are closed, inside and outside, by a pair of SS thin plates. The design for each vessel, with double layer walls, allows sectorized He-based leakage tests, to be progressively performed during the construction.

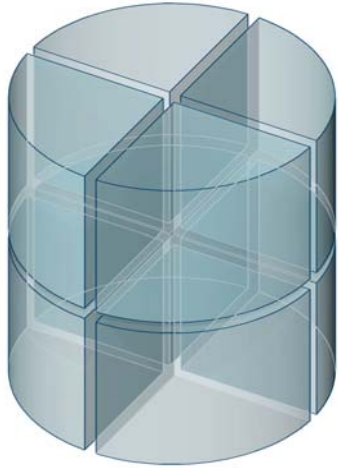

**Figure 3** – Liquid argon active volumes for a reinforced cryostat configuration.

For the inner vessel, the volume in between the double layers and inside the latticework structure is subdivided into horizontal ring regions. In normal operation, all ring regions are independently filled with $LN_2$ at a constant pressure of 3.6 bar (corresponding to a temperature of 90° K and to a LAr vapour pressure of 1.3 bar), for limiting the thermal gradient in the LAr and the consequent convection flows. The overpressure inside the wall thickness contributes to their rigidity and allows, by filling it with $LN_2$, for a preliminary low-temperature stability test, before the final filling with LAr.

This same principle can be adopted also for the outer vessel where the walls are pressurized with nitrogen vapour (eventually, in equilibrium with the $LN_2$ in the inner vessel jackets).

The two vessels (warm and cold) can be eventually linked together by a lattice structure made by insulating material (fibreglass rods eventually combined with thin wall SS tubes) to withstand the stress induced by the vacuum between them.

The above-described design, when extended to 10-100 kTon, needs further measures to increase its stiffness and mechanical stability. As first step-down compromise from the ideal single volume configuration, we can consider to split the single LAr volume into an array of instrumented volumes[e] (see Figure 3) and use the gaps between them to pass reinforcing horizontal and vertical girders connected to the cold vessel walls (this possibility is sketched also in Figure 2). We can take advantage of these dead space gaps for positioning inside them a distributed array of photo-multipliers required for the event $t_0$ generation. The spacing girders can be structured to be filled with $LN_2$ for increasing the temperature uniformity in the LAr.

## 4  The inside detector

One possible configuration of the inside time projection chamber is shown in Figure 4. The readout is made by two vertical wire chambers, each made by two wire planes with wires oriented at 0º and 90º. With a 3-mm wire pitch, the total number of wires and related electronic channels is 16640. Uniformity in wire lengths (~ 13 m) and electrical capacitances (~ 260 pF) results in a uniform *S/N* ratio in the drift volume imaging.

The necessity of minimizing the signal cable length (and the related parasitic capacitance[f]) suggests the following possible alternative solutions for which dedicated R&D activities are required:

*a)* Outer, warm traditional electronic chain combined with LAr-to-vacuum-to-outside cryogenic feedthroughs. The feedthroughs are aligned along one vertical and one horizontal side of the wire chamber frame.

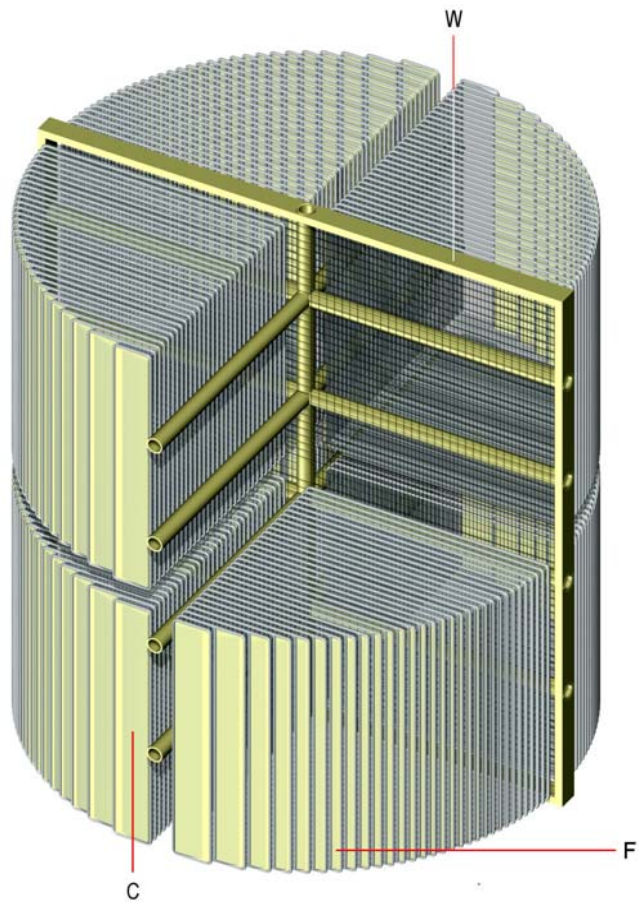

**Figure 4** – The inside detector, with one of the top quarters not shown. W: wire chambers; F: field-shaping electrodes; C: cathodes. The inner reinforcing cross is also indicated.

---

[e] Volumes with confined electric field for the drift
[f] Unitary cable capacitance: $C_c \sim 50$ pF/m

*b)* New read-out cryogenic electronic chain, including low-noise pre-amplifier, shaping amplifier, analog multiplexer and reduced number (1/32 or 1/64) of amplified signal cables.

## 5  The purification system

The well-known (see Ref. 1) LAr purification system in liquid phase needs improvements in the cryogenic pump design. The pump should operate in laminar regime to avoid microphonic noise and alteration of the LAr thermodynamic parameters. A prototype of such pump is being developed in the frame of the R&D activity of Ref. 4.

## 6  Conclusions

We have tried to underline the main guidelines and crucial points for the design of a multi-kTon LAr imaging detector. The extremely low $LN_2$ consumption, due to the double-wall vacuum insulation, and the reduced number of channels, due to the monolithic detector geometry and to the long drift, make the described example solution worthy of in-deep working out, in an engineering frame, for a realistic 2 kTon prototype. Essential requirements are a wise choice of materials and structure solutions aimed at achieving the UHP and UHV standards. Solutions coming from other application areas (LNG tanks) should be eventually revised according to these criteria. The geometry of this detector-cryostat complex is such to allow its scaling up to greater volumes in an economically convenient way. The described analysis suggests R&D activities on long path drift, on cryogenic adiabatic pumps, on cold feedthroughs and on a new medium scale integrated cryogenic electronic chain.